\documentstyle[psfig,prb,twocolumn,aps,floats]{revtex}
\begin{document}
\begin{widetext}

\title{The convergence of the ab-initio many-body expansion for the
cohesive energy of solid mercury}

\author{Beate Paulus, Krzysztof Rosciszewski$^*$ \\[0.5cm]
Max-Planck-Institut  f\"ur Physik komplexer Systeme, \\
N\"othnitzer Stra\ss e 38, D-01187 Dresden, Germany \\[0.5cm]
${}^{*}$ Institute of Physics, Jagellonian University,\\ Reymonta 4,
Pl 30-059 Krakow, Poland \\[1.cm]
Nicola Gaston, Peter Schwerdtfeger \\[0.5cm]
Theoretical Chemistry, Institute of Fundamental Sciences, Massey 
University (Albany Campus),
\\Private Bag 102904,
North Shore MSC, Auckland, New Zealand\\[1cm]
Hermann Stoll \\[0.5cm]
Institut f\"ur Theoretische Chemie, Universit\"at Stuttgart,\\
D-70550 Stuttgart, Germany
}

\maketitle

\begin{abstract}
A many-body expansion for mercury clusters of the
form
\begin{equation}
E =   \sum_{i<j  }\Delta \epsilon_{ij} + \sum_{i<j<k}\Delta
\epsilon_{ijk} + \ldots  \quad ,
\end{equation}
does not converge smoothly with increasing cluster size towards the
solid state.
Even for smaller cluster sizes (up to n=6), where van
der Waals forces still dominate,
one observes bad convergence
behaviour. For solid mercury the convergence of the
many-body
expansion can dramatically be improved by an incremental
procedure
within an embedded cluster approach. Here one adds the
coupled cluster many-body electron
correlation contributions of the
embedded cluster to the bulk HF energy. In this way we obtain
a
cohesive energy (not corrected for zero-point vibration) of 0.79 eV
in perfect agreement with
the experimental value.
\end{abstract}

\end{widetext}

\section{Introduction}
It is nowadays possible to calculate properties from first principles
for a wide range of materials.
This is mainly due to the success of density-functional theory (DFT)
\cite{DFT}.
Although the exact exchange-correlation functional is not known,
useful approximations are
available, the most widespread ones being the local-density
approximation (LDA) \cite{jones89} and the
generalized gradient approximation (GGA) \cite{perdew95}.
While DFT methods perform very well in a wide range of bonding
situations (metallic, covalent, ionic),
both LDA and GGA have serious problems describing weakly bound
systems such as van der Waals (vdW) crystals  \cite{engel00}.
Here, an alternative is to invoke wavefunction-based ab-initio
methods of quantum chemistry \cite{helgaker,werner_juelich}.
These methods allow for a proper account of the vdW interaction in
small clusters of atoms (dimers, trimers, ...),
and, inserting these data into a many-body expansion of the bulk
cohesive energy, it is possible to extract
from them reliable results for rare-gas crystals  \cite{roscis99}.

Mercury is a special case: starting from the vdW bound dimer
(dissociation energy: 0.05 eV \cite{zehnacker}),
there is a smooth transition from vdW clusters
(up to about 20 atoms) to covalent clusters and to the
metallic system (more than 400 atoms) \cite{haberland90,busani98},
finally leading to the metallic solid with its overlapping $sp$ bands
(cohesive energy per atom: 0.79 eV \cite{crc97}). Whereas
the weakly bound dimer can only be reliably described with highly
correlated wavefunction-based
methods (like the coupled-cluster (CC) approach) and extended
correlation consistent atomic basis sets
\cite{dolg96,schwerdtfeger01,peterson03}, one would expect that the
metallic solid should be well described within DFT.
Unfortunately, this seems to be not the case: various functionals
yield very different results
\cite{moyano02,schwerdt_priv}, from overbinding (by 30\% with LDA)
to severe underbinding (by a factor of nearly 3 with typical
GGA
functionals like PBE). Mixing in Hartree-Fock (HF) contributions
to form hybrid functionals like B3LYP does not help either: with pure
HF, the crystal energy at the equilibrium lattice constant
is repulsive by  about the same magnitude as the experimental value
is attractive \cite{paulus04};
the binding is entirely due to electron correlation.
On the other hand, it may not be surprising that a many-body
expansion derived from wavefunction-based
correlated cluster data (dimers, trimers, ...)
is of rather limited success also \cite{moyano02}, since the metallic
bonding of the solid has not much in common with the
vdW like behaviour
of the smaller clusters. In order to make such an expansion work, it
is necessary to introduce solid-state
information from the outset.

An attempt into this direction has been made very recently by two of
us \cite{paulus04}.
A method has been proposed for solid mercury, where the mean-field
(Hartree-Fock) part is calculated for the infinite
solid and only the correlation part is determined within a many-body
expansion using the wavefunction-based CC approach;
moreover, the CC calculations have been performed for finite, but
embedded clusters which mimic the confinement of the
electrons in the solid.
This way, very good agreement with the experimental cohesive energy
has been obtained.
In the present paper, we discuss this method in detail,
and we compare it with other variants
of many-body expansions, with and without embedding, to discuss the
convergence behaviour.
In the next section (Sec. II) we present our HF calculations. In Sec.
III we discuss different possible ways of setting up the
many-body expansion. In Sec. IV we construct the embedding.
After that (Secs. V to VIII) the individual contributions to the various
many-body expansions are presented.
   From these, the cohesive energy of mercury is evaluated in Sec. IX,
and conclusions follow in Sec. X.

\section{Periodic Hartree-Fock}

For calculating the HF contribution to the binding in bulk Hg
(rhombohedral structure, with equilibrium lattice constant $a_0 =
3.005$ {\AA}\,
and $\alpha = 70.53^{\rm o}$ \cite{crc97}),
we use the program package CRYSTAL98 \cite{crystal98}.
We changed the default parameters in order to obtain a
converged result for the HF binding energy, i.e.
we set integral thresholds to $\leq$10$^{-8}$ a.u. and convergence
criteria for the total energy and the orbital coefficients
to 10$^{-6}$ a.u. and 10$^{-5}$ a.u., respectively; our $k$-mesh
involved 9825 $k$-points in a Gilat net.
The chemically inactive [Kr]$4d^{10}4f^{14}$ core of the Hg atom is
simulated by an
energy-consistent scalar-relativistic pseudopotential (PP) \cite{andrae90}.
In CRYSTAL98, the Bloch functions of the solid are
generated using
local Gaussian-type (GTO) basis
functions. In principle, GTO sets optimized for atoms could be used;
however, diffuse functions describing atomic tails are not
necessary in the densely packed solid (and would in fact lead to
convergence difficulties in the SCF procedure if used nevertheless).
In our calculations we start from a $(7s7p6d1f)$ set of primitive
Gaussians\cite{peterson03};
we leave out the most diffuse $s$ and $p$ function and partially reoptimize
the next inner ones for the crystal (the resulting exponential parameters
are 0.126 for both $s$ and $p$); we next form fixed linear combinations
(contractions) of the first three $s$ and $p$ Gaussians and of the
first four $d$
Gaussians using the $5s$, $5p$, $5d$, and $6s$ atomic orbitals
as contraction coefficients; finally, we leave out the $f$ polarization
function because such functions are not too important for the
HF
cohesive energy of metals without an open $d$-shell.
The resulting contracted GTO (CGTO) basis set can be characterized as
$(6s6p6d)/[5s4p3d]$.
In order to provide an {\em a-posteriori} justification for the
truncation of the original atomic basis set,
we performed a free-atom calculation with the modified (crystal)
basis centered at the site of the atom {\em and}
at the 12 neighbouring atomic positions of the solid.
The resulting atomic HF energy is only about 1 mHartree
higher than that obtained  in a standard free-atom calculation with
the original uncontracted $(7s7p6d1f)$ basis set,
so that we can use the former value as a reference for calculating the
cohesive energy of the solid.

The calculated HF cohesive energy (binding energy per atom) of solid
mercury in the rhombohedral structure
is listed in Table \ref{cohen}. It is seen that there is no binding
at the HF level, at least not at the experimental
lattice constant of the crystal,
in spite of the metallic behaviour signalled by the overlap of the
bands derived from the 6$s$ and $6p$ atomic orbitals;
the HF repulsion is of about the same magnitude as the experimental
value of the cohesive energy.
There is no binding either at the HF
level for the dimer Hg$_2$ \cite{schwerdtfeger94}, and most likely
for all Hg$_n$ clusters up to
the solid state. This situation is
usually found for vdW systems only.

\section{Many-body expansions}

When setting up a many-body expansion of the form
\begin{equation}
\label{en}
     E = \sum_i             \epsilon_i +
                    \sum_{i<j  }\Delta \epsilon_{ij} +
                    \sum_{i<j<k}\Delta \epsilon_{ijk} + \ldots
     \quad ,
\end{equation}
for the crystal energy (or part of it), one first has to specify the
meaning of the $n$-body indices
$i,j,k,\ldots$. Both for a vdW crystal
and a metal, a numbering in terms of atoms seems to be natural. However,
only if local interactions prevail will $n$-body contributions for
distant atom pairs $i,j$, triples $i,j,k$, etc.\ decay fast enough to
make such an expansion useful.

In its simplest form, $E$ is the total crystal energy, the
$\epsilon_i$ are taken as the (total)
energies of the free atoms, the $\Delta \epsilon_{ij}$
are the non-additive parts of the total energies $\epsilon_{ij}$ for
(isolated) pairs of atoms $i,j$:
\begin{equation}
\label{twobody}
\Delta\epsilon_{ij}=\epsilon_{ij}-\epsilon_i - \epsilon_j ;
\end{equation}
similarly, the $\Delta \epsilon_{ijk}$ are non-additive parts of the
total energies
$\epsilon_{ijk}$ of trimers corrected for pair interactions:
\begin{equation}
\Delta\epsilon_{ijk}=\epsilon_{ijk}-(\Delta\epsilon_{ij}+\Delta\epsilon_{ik}+\Delta\epsilon_{jk})\\
-(\epsilon_{i}+\epsilon_{j}+\epsilon_{k}).
\end{equation}
Such an expansion has been shown to work very well for vdW crystals
like the rare gases \cite{roscis99}.
Here, the decay of the contributions is fast enough with distance
($\sim 1/r^6$ in the leading term
of the vdW pair interaction), the expansion
is dominated by pair contributions making higher terms almost negligible,
and the nature of the interaction changes very little when going from
small clusters (dimers, trimers, ...) to the infinite crystal.

It may be argued that it is just the last point which makes this type
of expansion less successful for mercury:
the metallic solid has not very much in common with
the vdW bound small clusters. However, we will see that even for
smaller
mercury clusters, where we would still expect
some vdW like behaviour,
the convergence of the many-body
expansion is not smooth at smaller distances.
An obvious possibility for improvement is to
treat the HF part of the total energy separately,
i.e., without many-body expansion, in a calculation for the bulk
solid like in Sec.\ II.
The many-body expansion for the remaining part of the
total energy (the correlation energy) is expected to be both less
problematic and more general, since i) electron correlation effects are known
to be more local than interactions at the independent-particle level, and ii)
the decay of electron-correlation contributions for distant pairs
of localized orbitals (or orbital groups) shows a vdW-like behaviour
not only in vdW crystals but also in ionic and covalently bonded systems.

Still, localized entities in ionic and covalently bonded solids may
be quite different from those in free (neutral)
atoms or small clusters.
Therefore, it may be vital to base the many-body expansion of the
correlation energy on
suitable localized orbitals or orbital groups as they
appear in the solid (e.g., atoms or ions modified by crystal surroundings,
bond orbitals, etc.); the necessary information can be taken directly
from solid-state calculations or from calculations for suitably
embedded clusters.
Such a type of expansion has been successfully applied by our
group to a wide range of ionic crystals and semiconductors,
cf.\ e.g.\ \cite{stoll92b,paulus96,doll95}.

A direct transfer of this approach to metallic systems is not
possible, however,
since localized orbitals become very long-range entities here and
a many-body expansion in terms of such orbitals cannot be expected to
have useful convergence characteristics. Moreover, it is currently
not clear how to treat electron correlation
within wave-function
based methods for metallic systems, since it becomes highly
multi-configurational in nature, and a full CI treatment is not
feasible.
In order to make the expansion still computationally feasible, we have
suggested recently \cite{paulus04} to start from suitable
model systems where long-range orbital tails are absent, and to allow
for delocalization only successively
in the course of the expansion; more specifically, when calculating
pair contributions for a given orbital (or orbital group)
combination ($i,j$), we allow for delocalization
$i \rightarrow j$ and $j \rightarrow i$, and similarly with the 3-body terms
we allow for delocalization over the triples of atoms, etc.
It is clear that the final result is not affected, only the
convergence properties of the many-body expansion are changed.
As an additional advantage, we can calculate individual terms of the
expansion from (suitably modelled/embedded)
finite clusters of reasonable size.
In the case of mercury, for example, we can force localization of the solid
by using a $s$-type atomic basis set for
describing the valence-electron system.
This way, delocalization due to $sp$-mixing is avoided,
but still each atom has its correct crystal surroundings concerning
the van der Waals interaction. When
determining a many-body contribution for a given set
of atoms, we can use the full basis for this set of atoms and thus
successively allow for metallic delocalization.

\section{Many-body increments for embedded clusters}

As already mentioned in the previous section,
we want to use finite cluster models for  embedded $n$-tuples of
atoms, in order to calculate
individual terms ($n$-body increments) of the many-body expansion of mercury.
These cluster models have to meet the following requirements:
i) their geometry should reflect the experimental geometry of a
suitably chosen section
of the Hg crystal, in order to mimic the influence of the bulk surroundings
on the $n$ inner atoms to be correlated; ii) their electronic
structure should simulate
that of a hypothetical Hg crystal with the highest occupied band
derived from atomic $6s$ states only,
whose Wannier orbitals are well localized and well transferable,
without significant finite-size and
surface effects, to the cluster models in question.

According to requirement i), the geometries of the embedded clusters
were generated as follows. The rhombohedral structure
of the infinite crystal can be viewed as a central atom surrounded by
atom shells
of various size.
The first shell contains 12 atoms,
6 of them at distance $a_0$ (=3.005  {\AA}) and 6 at 1.155 $a_0$; the
next shells
contain 6 atoms at
distance 1.528 $a_0$, 6 atoms at 1.633
$a_0$ another
24 atoms up to a distance of 2 $a_0$.
This already defines various levels of embedding for a single atom
when calculating the one-body term
of the many-body expansion. For calculating a two-body term between
neighbouring atoms, we include all atoms in the embedding
which are in the first shell of one of the two
atoms to be correlated.
If the two atoms to be correlated are more distant, we additionally
select for the embedding
all atoms lying within a cylinder of the shell radius around the
connection line.
Since the number of embedding atoms is large, we do not use here the
small-core (20-valence-electron) pseudopotential
mentioned in Sec. 2, but rather a  2-valence-electron
scalar relativistic  pseudopotential\cite{kuechle91}
which simulates the Hg $5s^25p^65d^{10}$ shells
within the atomic core. Thus, only the $6s, 6p$ and higher atomic
shells are explicitly treated;
truncating the corresponding optimized $(4s4p1d)/[2s2p1d]$ valence
basis set to $(4s)/[2s]$,
we satisfy requirement ii).

For calculating the individual $n$-body increments in the embedded
clusters, we have to treat
the orbitals of the $n$ atoms involved (which are in the center of
the embedded cluster)
as accurately as possible.
We equip these atoms
with the small-core pseudopotential, i.e.,
we explicitly treat the outer-core
$5spd$ shells in the valence space; furthermore, we invoke the
(unmodified) primitive
$(7s7p6d1f)$, $(10s9p7p2f1g)$, and $(12s12p9d3f2g1h)$ basis sets of
Ref.\ \cite{peterson03}.
Two different contraction patterns of these primitive basis sets are
considered in the following.
In the first step of our calculation, the aim is to define the
localized orbitals
corresponding to the hypothetical mercury mentioned above. Here, we
choose contractions
which are closely analogous to that for the embedding atoms: the
outer-core $5spd$ orbitals and the
valence $6s$ orbital are fully contracted using the orbital
coefficients of the free atoms, and the
most diffuse $s$ function of the primitive sets are added to provide
more flexibility within the $s$ space;
this leads to a $[3s1p1d]$ set.
Since valence $p$ functions are not represented in these basis sets,
delocalization of the orbitals is still avoided. Thus, a unitary
transformation of
the occupied canonical orbitals according to the criterion of Foster
and Boys \cite{foster60} yields well
localized orbitals on the individual atoms, which can be separated
into embedding orbitals
and orbitals to be correlated.

In the next step, we improve the description of the atoms to be
correlated while keeping frozen
the localized orbitals which can be attributed to the atoms of the
embedding region.
The basis sets of the former atoms are enlarged by successively
decontracting the
above-mentioned basis sets with respect to the most diffuse exponents.
Specifically we consider a
$(7s7p6d1f)/[7s6p5d1f]$ set (basis A),
a  $(10s9p7d2f1g)/[8s7p6d2f1g]$ set (basis B),
and a $(12s12p9d3f2g1h)/[9s8p7d3f2g1h]$ set (basis C).
Using these basis sets, we recalculate the integrals and  reoptimize the
orbitals of the atoms  to be correlated, in a
HF calculation. This provides us with orbitals which are still fairly
local but are more or less delocalized over
the atoms $i, j, .. $ to be correlated. Finally, on top of this HF
calculation and still keeping
frozen the localized orbitals of the embedding region,
we introduce electron correlation by performing a coupled-cluster
calculation with single and
double excitations and perturbative treatment of the triples (CCSD(T)).
All these calculations are performed using
the MOLPRO suite of ab-initio programs
\cite{molpro2002,hampel92,deegan94}. Note that only the
correlation-energy piece, $\epsilon_{i,j,..}$,
of the last calculation enters the many-body expansion of Sec. III
(eqs. 1 - 3).

Test calculations have been performed checking the influence
of geometry (number of shells with embedding atoms) and basis-set
description of the embedding region on the
calculated $n$-body increments.
For the one-body increment we checked clusters with
up to 7 shells, for the
nearest-neighbour two-body increment clusters with up to 4 shells.
All calculated one-body and two-body increments differ by at most by
0.1mHartree, respectively.
On the basis of these test calculations and in order to avoid
excessive computational effort, we feel justified to restrict
embedding to the first shell, for the calculations of the following sections,
and to describe  the embedding atoms with the large-core
pseudopotential and the corresponding
$(4s)/[2s]$ basis set.

\section{One-body contribution to cohesion}

It is only for the many-body expansion with embedding that the
one-body increment
can contribute to the binding. We define the cohesive contribution of
the one-body
increment as the difference between the correlation energy of the
embedded atom $\epsilon_i$ and that of the free atom
$E_{\rm free}^{\rm corr}$.
\begin{equation}
\epsilon_i^{\rm coh}= \epsilon_i - E_{\rm free}^{\rm corr}
\end{equation}
The free atom is calculated with the same basis as the embedded atom,
but in order to minimize basis-set superposition effects
we add one shell of ghost atoms (i.e., atoms with zero charge carrying
(4s)/[2s] basis sets) at the 12 nearest-neighbour sites of the mercury
lattice. The difference between the correlation energy with this basis set
and that with the uncontracted basis set of the
free atom is less than 1 mHartree.

In Table \ref{onebody}, the one-body contribution to the binding is listed for
different basis sets, and
different number of correlated atomic orbital shells.
All contributions are repulsive; the embedded atom has a
smaller (absolute) correlation energy than the free atom.
This effect is due to the crystal cage effect leading to a more
compact $6s^2$ shell
but also to increased excitation energies.
As can be anticipated the dominant correlation
contributions to $\epsilon_i^{\rm coh}$
come from the $6s^2$ shell. Intra-shell contributions from the
outer-core shells are
also repulsive, but significantly smaller; moreover the latter
contributions are nearly
compensated by the $5d-6s$ inter-shell
correlation  which becomes increasingly attractive with increasing
compactness of the 'in-crystal' atoms.
Interestingly, although $5sp-5d6s$ inter-shell effects are small in
magnitude they can be seen
to become repulsive again. Finally, basis effects are small: adding diffuse
functions (as detailed in Sect.\ 6), or going from basis set B to C
changes the total $\epsilon_i^{\rm coh}$ by not more than 0.1 mHartree.

\section{Two-body contributions}

\begin{figure}
\begin{center}
\psfig{figure=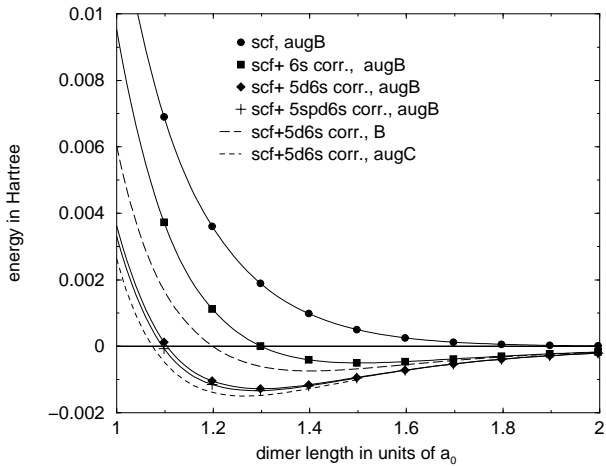,width=8cm,angle=-90}
\end{center}
\caption{\label{dimer}
The potential curve of the Hg dimer is plotted for different basis
sets (B, augB, augC)
at the Hartree-Fock (scf) and correlated levels (corr.). In the
latter case, the active
space of correlated orbitals is indicated ($6s$, $5d6s$,
$5spd6s$).
$a_0 =  3.005$ {\AA}}
\end{figure}

In a first step, we checked our basis sets for
the free dimer where experimental values and data from accurate
calculations are available in
literature\cite{dolg96,schwerdtfeger01,peterson03}.
Since the vdW binding is very weak (of the order of 1 mHartree) and
relies on an accurate description
of the orbital tails, we added diffuse functions to our basis sets;
more specifically,
one diffuse function for each $l$ value was added in an even-tempered
way (the resulting basis sets
being designated as augA, augB, augC).
In Fig.\ref{dimer}, the Hg$_2$ potential curve is plotted for basis sets B,
augB, and augC,
for various choices of the active space of correlated orbitals;
all curves contain counterpoise corrections \cite{boysbernardi}
for (approximately) removing basis-set superposition errors.
The HF curve is purely repulsive. Correlating the $6s^2$ shell leads
to binding, but the dissociation energy $D_e$ is too small by more
than a factor of 3, with respect to the experimental value, and the
bond length $R_e$ is too long by more than 0.6 {\AA}.
Only with correlating the outer-core $5d^{10}$ shell is it possible
to obtain reasonable results; the influence of
the underlying $5s^25p^6$ shells, on the other hand, is only
marginal. More important than the latter correlation
effect is the role of the basis set; especially prominent is the role
of the diffuse functions,
as shown by comparing the B and augB curves: $D_e$ is enhanced by
nearly a factor of 2, $R_e$ is shortened by $\sim$0.3 {\AA};
going from basis augB to augC leads to  further improvement and
yields the following
values for the spectroscopic constants (experimental values in
parentheses): 3.80 (3.69$\pm$0.01) {\AA}\ for $R_e$, 17.9
(19.6$\pm$0.3)
cm$^{-1}$ for the harmonic wavenumber $\omega_e$, and 328
(380$\pm$15) cm$^{-1}$ for  $D_e$.

\begin{table}
\begin{tabular}{|r|c|c|c|}
active space & augC&augB&B\\
\hline
$6s^2$&             +0.004154&+0.004155&+0.004230\\
$5d^{10}6s^2$&      +0.004436&+0.004582&+0.004260\\
$5s^2p^6d^{10}6s^2$&+0.005041&+0.005104&+0.005089\\
\end{tabular}
\caption{\label{onebody} The one-body contribution to the cohesive
energy $\epsilon_i - E_{\rm corr}^{\rm free}$
of embedded clusters (in Hartree)
is calculated at the CCSD(T) level with
different basis sets (B, augB, augC), and different
active spaces included in the correlation treatment. For more
details, cf.\ text.
}
\end{table}

We now use the above free-dimer data for the
many-body expansion of solid mercury.
The relevant increments are given in Table \ref{2body}. It is
seen that for the nearest-neighbour increment ($d = a_0$) $d$-shell
correlation enhances the valence ($6s^2$) correlation effect by more
than a factor of 2, but this factor gradually diminishes to
$\sim 10 - 20\%$ for $d=1.9 a_0$. The inclusion of diffuse basis
functions is important; the $\Delta \epsilon_{ij}$ become
more attractive by 20 - 30 \% when going from basis set B to augB
(and by another 5 - 10 \% when going from aug-B to augC).
While all correlation-energy increments are attractive, the
nearest-neighbour increment becomes repulsive when including
the Hartree-Fock energy into the many-body expansion. Since the
Hartree-Fock closed-shell Pauli repulsion is a short-range
effect, already the 2nd-nearest-neighbour increment $d=1.15 a_0$
remains slightly attractive, and the increments with $d > 1.6 a_0$
are hardly affected at all.

Correlation-energy increments $\Delta \epsilon_{ij}$ for embedded
clusters are also listed in Table \ref{2body},
again for different basis sets
(B, augB, augC) and correlated active spaces ($6s^2$,
$5d^{10}6s^2$).
The basis-set dependence
turns out to be roughly the same as for the free dimers
(except for the fact that the diffuse functions are less important
because they are already simulated by the basis
sets of the embedding atoms).
However, the nearest-neighbour increment is larger by $\sim$15\% as
compared to the free-dimer case, while changes into the
opposite direction occur for the increments between more distant
atoms. These findings may be rationalized
in terms of the crystal cage effect
leading to an increased overlap between nearest neighbours (the
embedding forcing the electrons
to stay nearer to the atoms and to the bond between them) but a
reduced 'in-crystal' polarizability of the atomic entities.
Again, the contribution of the outer-core $5d$ shell is very
important for the nearest-neighbour increment.
As for the free dimers, the overall decay of the increments with
distance is fairly rapid. Moreover,
from a distance
around 1.5$a_0$ on the basis set effects are very small on an
absolute scale and hence can be neglected; there,
basis set B is sufficient.
For basis B, we have performed the calculation of the 2-body
increments up to a distance of
3.0 $a_0$, where the increment is less than 10 $\mu$Hartree. We have
fitted a vdW expression to the increments of the region
from 1.5 $a_0$ to 2.8 $a_0$. The resulting value for $C_6$, 297
atomic units, is comparable with data from the literature
\cite{schwerdtfeger01,hartke02,kunz96} where the van der
Waals constant for the free dimer was determined.

\section{Three-body contributions}

As found by Schwerdtfeger and co-workers \cite{moyano02}, three-body
terms are very
important for mercury clusters and solid mercury, especially the short-range
contributions. We have calculated, therefore, all 3-body
increments where at least 2 distances are within the first-neighbour shell
(i.e., smaller than 1.16 $a_0$), using basis sets B and augB.
The results are given in Table \ref{3body}.

For the free trimers, the largest contributions arise for the compact
geometries;
for the simplest type of many-body expansion (total-energy expansion),
these contributions are of the same magnitude as the largest two-body
increments;
since the number of 3-body increments (including weight factors) is
larger than that
of the two-body ones, the two-body repulsion is thus overcompensated by
three-body attraction. The three-body attraction
for Hg$_n$ clusters is in sharp
contrast to the rare gas elements, which
has the consequence that the
Hg$_3$ distance  becomes shorter compared to Hg$_2$ \cite{moyano02}.

The situation is different when the many-body expansion is performed for the
correlation energy only. Here, the two-body terms are attractive, and
this attraction is
weakened by (mostly repulsive) three-body increments;
the ratio of the largest two-body and three-body increments,
respectively, is $\sim$6:1.
The largest three-body increments still arise for compact geometries,
and are by a factor of 2 smaller than in the total-energy case.
The basis-set dependence turns out to be less critical for the
three-body increments than for the two-body ones: adding diffuse functions
to basis set B (B $\rightarrow$ augB) changes the individual
increments by mostly less than 0.1 mHartree.
Thus, we decided to leave out the diffuse functions for the embedded
clusters; here,
the role of the latter functions is simulated by the basis sets of
the embedding atoms,
leading to a reduced basis-set dependency already for the two-body
increments (cf.\
Table \ref{2body}).

Looking at the results for the embedded trimers now,
we see that the convergence behaviour of the many-body expansion is
significantly
improved once more: the largest three-body correlation-energy
increment is 4 times smaller
than for the free-trimer case, and $\sim$20 times smaller than the
largest (embedded) two-body increment.
The embedded three-body terms are mostly attractive (i.e., of the
same sign as the
embedded two-body terms); valence ($6s^2$) correlation dominates,
except in cases where at least one distance
of the (embedded) trimer is the shortest possible (nearest-neighbour) one --
there, $6s-5d$ inter-shell correlation is of a magnitude nearly
comparable to that of $6s^2$ correlation.
For the nearly equidistant
triangle the $5d$ correlation is attractive, but for the linear arrangement
of 3 atoms it is repulsive. This behaviour can be rationalized in terms of the
Axilrod-Teller dipole-dipole-dipole interaction
\cite{axilrod43}.
The valence-only values show just the reverse dependence on the angle.
By far the largest (total) contribution (-0.4 mHartree) is the
attractive one for the linear arrangement;
this might be connected to the fact that static polarization of the
central atom is suppressed there
and is (partially) compensated by correlation.
For the linear trimer, we checked the basis-set dependency: going
from basis set B to C
changes the increment from -0.4 to -0.5 mHartree.

\section{Four- and higher body contributions}

\begin{table}
\begin{tabular}{|l|rr|r|}
    & \multicolumn{2}{c |}{free}&embedded \\
&scf+$5d^{10}6s^2$&$5d^{10}6s^2$&$5d^{10}6s^2$\\
    \hline
tetrahedral-like&  +6588.4   &  -2373.3  &  -5.9      \\
planar, Y-like  &  +742.8    &  -726.4   &  +107.6   \\
planar, L-like  &  -73.5     &  -40.3    &  +93.0   \\
linear          &  +321.8    &  -54.3    &  +170.1  \\
planar,rhombus  &  -1090.7   &  -820.4   &  -20.0     \\
\end{tabular}
\caption{\label{4body} The four-body increments, $\Delta
\epsilon_{ijkl}$ (in $\mu$Hartree) for  compact 4-atom clusters of the
rhombohedral structure are calculated at the CCSD(T) level
with basis A. For the free and embedded clusters the
correlation-energy increments for the $5d^{10}6s^2$
active space
are listed, for the free clusters also the total-energy contribution
is given.
}
\end{table}

For the four-body contributions, we selected 5 compact geometries, the
tetra\-hedron-like one,  a linear one, and three planar geometries (a
rhombus, and two geometries with
3 atoms in line and the fourth one connected to the first atom
(L-like) or to the middle atom (Y-like)).
The results are listed in Table \ref{4body}.
They were obtained without diffuse functions, and basis set A was
used instead of B
in order to reduce the computational effort. We checked, however, for
the linear geometry that
basis B would change this increment by just about 20\%.
For the embedded case, the increments are smaller in magnitude than
the 3-body terms, their sum is repulsive.
The linear geometry has the
largest contribution (0.2 mHartree). This is totally different for
the free tetramers:
As for the corresponding 3-body terms, the compact ones yield the
largest absolute contribution both at the HF and the
correlated level (6.6 mHartree in total, -2.4 mHartree for the
correlation piece).
Considering the magnitude of the  four-body increments, we see
that the values do not decrease with respect to those for the trimers.
Thus, the convergence of a many-body expansion based on non-embedded
Hg clusters appears to be questionable,
even if it is done for the correlation energy only.

\begin{figure}
\begin{center}
\psfig{figure=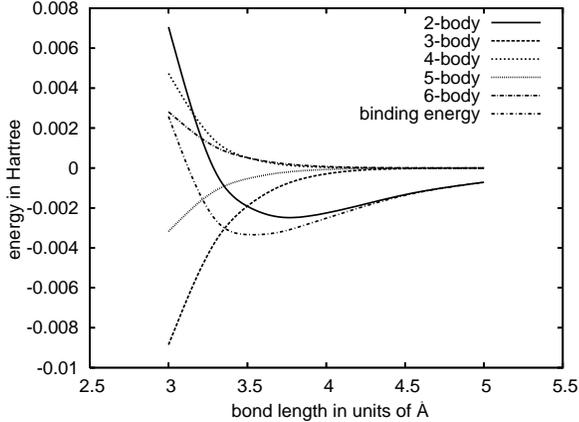,width=8cm,angle=0}
\end{center}
\caption{\label{dist}
Individual 2-,
3-, 4-, 5- and 6-body contributions (in Hartree)  to the
dissociation
energy per atom  for Hg$_6$ at different Hg-Hg bond
distances.}

\end{figure}

To make this
point even more transparent, we calculated all $n$-body contributions
up to $n=$6 for octahedral Hg$_6$.
Since coupled cluster calculations
are computationally too demanding for this size, we carried out
second-order
many-body perturbation theory (MBPT2) calculations for
the correlation energy using a somewhat reduced correlation
consistent
$(6s5p4d1f)/[4s4p3d1f]$ basis set (basis D) which
minimizes the basis set superposition error (BSSE).
The $n$-body contributions are shown in
Fig.\ref{dist} as a
function of the Hg-Hg distance in Hg$_6$ kept in
$O_h$ symmetry.
At short bond distances (3.0 \AA) the two-body part becomes repulsive 
as expected.
However the 3-body part remains attractive, giving -343\% of the 
total (non-binding) energy in contrast to
the +274\% due to the 2-body interaction.  The higher body terms 
contribute 183\%, -123\%, and 109\% for the
4-, 5-, and 6-body contributions respectively (a negative sign 
indicates binding here where the total energy is positive in contrast 
to the case below). At such short distances the HOMO/LUMO gap is 
still 

\begin{figure}
\begin{center}
\psfig{figure=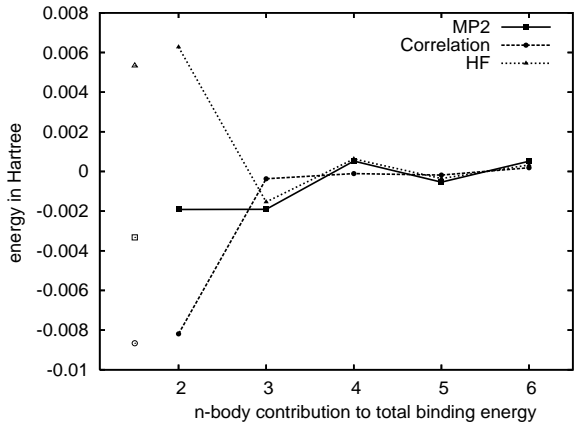,width=8cm,angle=0}
\end{center}
\caption{\label{body}Individual 2-, 3-, 4-, 5- and 6-body
contributions (in Hartree) to
the
dissociation energy per atom for Hg$_6$ at the equilibrium MBPT2
geometry.
The disconnected symbols are the total energy contributions summed
over all n-body interactions.}
\end{figure}

large. Hence, the bad convergence 
of the many-body expansion 
is not directly related to the change to the 
metallic state. Above 3 
{\AA} 
the 2-body part becomes attractive, and the contributions of 
each many-body term relative to the
binding energy of the 6-atom cluster decrease.   While at short 
distances the 3-body term is the contribution
of highest magnitude, above 3.5 {\AA} the 2-body term becomes 
increasingly more important.  However even at 4 {\AA} the 3-body term 
accounts for +12\% of the binding energy, followed by 4-, 5-, and 
6-body contributions of -1.4\%, 2.9\%, and -2.9\%.
While these contributions are not entirely negligible, it is clear 
that at longer distances the many-body expansion begins to converge. 
According to an earlier study we expect mostly
vdW behaviour at distances down to
5 {\AA} \cite{schwerdtfeger94}.  Hence, the bad convergence of the 
many-body expansion
is related to the fact that the bond distance 
shortens dramatically from 
the more vdW type system of Hg$_2$ (3.69 
\AA) to the solid state (3.0 \AA).

We also investigated the convergence behaviour of the energy
needed to remove one Hg atom from the cluster, which in the
high $n$ limit also converges towards the cohesive energy. Again, we
do not see a great improvement in
the convergence behaviour of the $n$-body expansion.

Fig. \ref{body} shows the HF,
correlation and total $n$-body contributions for Hg$_6$ at the
optimized MBPT2 equilibrium distance of 3.499 {\AA}.
The oscillating behaviour of the $n$-body terms is clearly seen.
The 2- and 3-body contributions are almost exactly equal, 58\% and 57\% of the
total binding energy. The 4-, 5-, and 6-body contributions are all important at
-16\%, +16\% and -15\% respectively. Fig. \ref{body} also
shows the sum over all $n$-body
contributions (detached symbols shown at left), which
demonstrates that while the $n$-body expansion is starting to converge, the
higher $n$-body contributions are still not negligible at this size.
Hence, the expansion shown in eq.1 is of limited use for the total 
energy of free
clusters.  However, the correlation energy shows a much better 
convergence behavior at 
3.499 {\AA}, Fig. \ref{body}. Even at 3.0 
{\AA} the 2-body correlation energy is the
major term with 94.3\%, 
the higher-order terms becoming much smaller with
the 3-body 2.0\%, 
the 4-body 4.7\%, the 5-body 1.2\%, and the 6-body -2.3\%.
Obviously, 
this explains the nice convergence behavior in the solid state 
calculations
for the correlation part.

\section{Results for the cohesive energy of bulk mercury}

\begin{table}
\begin{tabular}{|l|r|r|r|}
& CRY& CRY&  MB\\
& emb& free&free\\
\hline
HF&+36.2 &+36.2&\\
\hline
1-body, basis B &+4.2 &&\\
1-body, basis augB&+4.6&&\\
1-body, basis augC&+4.4&&\\
\hline
2-body, basis B &-50.0&-44.7&+12.7\\
2-body, basis augB &-57.5$^*$&-56.7&-0.1\\
2-body, basis augC &-61.2$^*$&-61.3&-4.8\\
\hline
3-body(1st), basis B &-9.2&+20.0&-39.5\\
3-body(1st), basis augB &---&+19.1&-40.4\\
\hline
3-body(2nd), basis B &-2.4&+2.8&-2.9\\
3-body(2nd), basis augB &---&+0.8&-4.0\\
\hline
4-body, basis A&+3.0&-16.6&+18.9\\
\hline
total, best available basis& -29.2&-21.8&-30.3
\end{tabular}
\caption{\label{cohen} Contributions to the cohesive energy of solid
mercury (in mHartree) are listed
for different many-body expansions (including the Hartree-Fock energy
in the many-body
expansion (MB) or using the crystal HF energy (CRY), calculating the
correlation
energy increments from free (free) or embedded clusters (emb)) and
different basis sets (A, B,
augB, augC).
For the 1-body term the contribution of the free atom is subtracted.
The 2-body increments are summed up to a distance of 2.52 $a_0$. (For
the augmented basis sets, the
$^*$ means that only increments up to a distance of 1.92 $a_0$ are
calculated with the  basis set indicated,
while basis set B is used for the rest.)
The sum of the 3-body increments for the first shell is taken over
the increments with
2 distances smaller 1.16 $a_0$, for the second shell over increments with
2 distances smaller than 1.63 $a_0$.
The 4-body contributions is the sum over 5
compact clusters.
The experimental value is -29 mHartree \cite{crc97}.}
\end{table}

In Table \ref{cohen} we list the various contributions to the cohesive energy
of solid mercury applying different types of many-body expansion.
Looking first at results
for the many-body expansion
of the correlation energy derived from embedded-cluster data, we see
that the 2-body term is the dominant one, the 3-body contribution of
the first shell
being smaller by a factor of 5. There is a further decrease
by about a factor
of 3 when going to the 3-body terms of the second shell or to the
most important 4-body terms.
Summing up all contributions calculated with the largest available basis set
(Table \ref{cohen}) we obtain -29.2 mHartree.
The cohesive energy is
therefore 0.79 eV, in very good agreement with the experimental value.
But the approximations applied, especially the finite one-particle 
basis set and the neglect of
further 4-body terms yield an error bar of about 10\% for the calculated value of the cohesive energy.

Replacing the many-body expansion for the
correlation energy of the embedded
clusters by a corresponding expansion for free clusters, leads only
to a moderate decrease of the (absolute value of
the) binding energy. 
However, the 3-body and 4-body
terms are of comparable magnitude
(and of the same oder of magnitude as the experimental
cohesive energy) indicating that
the result is not properly converged.
Similar convergence problems are encountered with a
many-body expansion of the total energy (HF + correlation)
for free Hg clusters. Again, the sum of the increments considered in
this paper would lead to more or less reasonable
results, but the 3-body contribution is
now more than three times larger than the 2-body one (while the factor
was about 1/3 for the correlation-only
many-body expansion for free clusters).
Thus, the many-body correlation-energy expansion for
embedded clusters is found to provide
more reliable results. It is to be seen how
well this new type of expansion, which
has been applied to metals for the first time in the present and a
companion \cite{paulus04} paper,
works for other properties of mercury like lattice constants and the 
bulk modulus.

\section{Conclusions and Outlook}

We have demonstrated that the cohesive energy of mercury can be
obtained
with good accuracy from solid state HF calculations
adding
$n$-body correlation calculations for embedded mercury
clusters. However, the
simulation of free mercury clusters where
surface effects are dominant will represent
a considerable challenge
for future quantum theoretical investigations.
First, the n-body
expansion only converges for free clusters above a critical bond-length where
van der Waals bonding becomes dominant.
The bad convergence of the $n$-body expansion for mercury
can subsequently lead to a
bad convergence of the vdW
equation which contains the well known virial coefficients.
Hence,
the accurate simulation of gaseous or liquid mercury
is currently
a formidable task. Second, DFT does not produce reliable results
for such
clusters and an improvement for describing both low and high $n$ limits
is required,
that is the vdW system for smaller clusters,
and metallic state for the solid. Third, single-reference wavefunction
based methods such as CCSD(T)
will fail
as the cluster
becomes closer to the metallic state with increasing
size.
A possible way out of this dilemma can be a multi-reference incremental scheme,
as it was successfully applied for the one-dimensional model system
lithium\cite{paulus03a}.
Combining this method with the embedding proposed here for metallic systems
could provide a quantum chemical method for metals like barium, where the
metallicity is stronger than in mercury.

\section*{Acknowledgements}
Support by the Marsden Fund (Wellington) is gratefully acknowledged.
N.~ G.~ acknowledges support from FRST (Wellington).

\newpage
\begin{widetext}

\begin{table}
\begin{tabular}{|r|rr|rrr|rrrr|}
    & \multicolumn{5}{c|}{free} &  \multicolumn{4}{c|}{embedded}\\
&\multicolumn{2}{c|}{scf+$5d^{10}6s^2$}&$6s^2$ &
\multicolumn{2}{c|}{$5d^{10}6s^2$}& $6s^2$ & \multicolumn{3}{c
|}{$5d^{10}6s^2$}\\
$ d_{ij}/a_0$ & augC & augB& augC&augC & augB& augC& augC&augB&B \\
\hline
1.000  & +2579  & +3619& -4030 & -10759&-9805&-4592&-12033&-11243&  -9842\\
1.16  & -1140  &  -701& -2858 & -5909 &-5476&-2706&-5636 &-5310 &  -4545 \\
1.53  & -927   &  -867& -935  & -1336 &-1281&-594 &-928  &-891  & -754   \\
1.63  & -679   &  -643& -652  & -878  &-844 &-462 &-674  &-650  & -563   \\
1.92 &  -277   &  -266& -248  &  -303 &-292 & -236&-285  &-274  & -210   \\
\end{tabular}
\caption{\label{2body} The two-body increments $\Delta
\epsilon_{ij}$ (in $\mu$Hartree) for different distances $d_{ij}$
occurring in the
rhombohedral structure are calculated for the free and embedded
dimers at the CCSD(T) level
using different active spaces of
correlated orbitals ($6s^2$, $5d^{10}6s^2$) and different basis sets
(B, augB, augC).
The values in the columns `scf+$5d^{10}6s^2$' include the Hartree-Fock
energy within the many-body expansion, cf.\ Sect. 3.
}
\end{table}

\begin{table}
\begin{tabular}{|ccc|rrr|rr|}
& & & \multicolumn{3}{c|} {free}  &  \multicolumn{2}{c |}{embedded } \\
& & & scf+$5d^{10}6s^2$&\multicolumn{2}{c|}{$5d^{10}6s^2$
}&$6s^2$&$5d^{10}6s^2$\\
$d_{12}$&$d_{13}$&$d_{23}$&augB&augB&B&B&B \\
\hline
     1.00 &   1.00 &   1.16 & -3012.8 & +1704.1& +1608.6&+83.4  &   -91.5 \\
     1.00 &   1.00 &   1.63 & -914.6  & +423.5 & +460.4 &-209.9&    -329.0\\
     1.00 &   1.00 &   2.00 & -174.2  & -419.9 & -249.6 &-579.4&    -423.2\\
     1.00 &   1.16 &   1.53 & -696.7  & +586.1 & +568.9 &-359.5&    -124.4\\
      1.00&   1.16 &   1.92 & -302.7  & -46.3  & +7.5   &-206.7 &   -245.7\\
      1.16&   1.16 &   1.16 & -1264.7 & +1035.2& +973.0 &+86.6 &     +94.3\\
      1.16&   1.16 &   2.00 & -163.7  & -21.7  & +3.7   &-102.7&    -125.9\\
      1.16&   1.16 &   2.31 & -147.3  & -239.7 & -185.8 &-179.2&    -151.4\\
\end{tabular}
\caption{\label{3body} The three-body increments $\Delta
\epsilon_{ijk}$ (in $\mu$Hartree) for compact 3-atom clusters of the
rhombohedral structure are calculated at the CCSD(T) level
with different basis sets (B and augB),
and different correlated orbital spaces ($6s^2$, $5d^{10}6s^2$).
The $d_{ij}$ are interatomic distances (in units of $a_0$).
Results are given for trimers with and without embedding (embedded, free).
The values in the column `scf+$5d^{10}6s^2$'
include the Hartree-Fock energy within the many-body expansion, cf.\ Sect. 3.}
\end{table}

\end{widetext}
\end{document}